\documentclass[prb,groupedaddress,showpacs,twocolumn]{revtex4}
\usepackage{epsfig}
\def\figwidth{8cm}
\begin{document}
\title{Meissner effect in a bosonic ladder}
\author{E. Orignac}
\email{orignac@lpt.ens.fr}
\affiliation{LPTENS CNRS UMR 8549 24, Rue
Lhomond 75231 Paris Cedex 05, France}
\author{T. Giamarchi}
\email{giam@lps.u-psud.fr}
\affiliation{Laboratoire de Physique
des Solides, CNRS-UMR 8502, UPS B\^at. 510, 91405 Orsay France}

\date{\today}
\begin{abstract}
We investigate the effect of a magnetic field on a bosonic ladder. We show that
such a system leads to the one dimensional equivalent of a vortex lattice in a
superconductor. We investigate the physical properties of the vortex phase,
such as vortex density and vortex correlation functions
and show that magnetization has plateaus for some commensurate values of the magnetic field.
The lowest plateau corresponds to a true Meissner to vortex transition at a
critical field $H_{c1}$ that exists
although the system has no long range superconducting order.
Implications for experimental realizations such as Josephson junction arrays are discussed.
\end{abstract}

\pacs{{05.30.Jp}, {71.10.Pm}, {74.50.+r}}

\maketitle

The effect of a magnetic field on interacting particles is a long
standing problem. A spectacular case is provided by type II superconductor,
in which the magnetic field is totally expelled below $H_{c_1}$,
whereas a vortex state exists for $H>H_{c_1}$.
This behavior, however, is obtained from the Landau-Ginzburg equation, and it is important
to know what happens when interactions and fluctuations
have more drastic effects, such as in one dimensional systems.
Indeed, in a one dimensional conductor at $T=0$, although there is no long range
order, superconductivity in
the sense of infinite d.c. conductivity can nevertheless be
present \cite{mikeska_supra_1d}. For a  one
dimensional chain, there is no orbital effect, and this question
is not relevant. However, a system made of a finite number of
coupled chains (a ladder), is still  one dimensional  (no
long range order can exist) but orbital effect of the magnetic field
is  present, opening the possibility of such a transition.

Beyond its own theoretical
interest the investigation of the effect of a magnetic field
on ladder systems is also of direct experimental relevance,
due to the various realizations of such ladders
\cite{dagotto_2ch_review,bockrath_luttinger_nanotubes,fazio_josephson_review}.
Fermionic ladders can be superconducting both from
attractive (s-wave) and repulsive interactions (d-wave).
In the attractive case, the system is close to standard
superconductors where pairs of fermions can hop from one chain to the
other leading to a Josephson coupling, provided the applied magnetic field is
smaller that the spin gap. The system
can thus be described as a bosonic ladder.
Josephson junction arrays
\cite{vanoudenaarden_josephson_mott,vanoudenaarden_josephson_localization}
provide also a very direct realization of such a bosonic ladder
\cite{bradley_josephson_chain,glazman_josephson_1d}
and are thus the prime candidates to
observe these effects. This problem of Josephson ladders has been
investigated previously in the classical
\cite{denniston_classical_ladder} and quantum limit both
analytically
\cite{kardar_josephson_ladder,granato_josephson_ladder}
and numerically \cite{nishiyama_josephson_ladder} in the high
field limit of half a flux quantum per plaquette for one
dimensional situations. For the Josephson two leg ladder, it has been shown
\cite{kardar_josephson_ladder,granato_josephson_ladder} that a true transition exists between
a commensurate and a vortex phase, however the detailed behavior
of the vortex phase and the effects of commensurability of the magnetic field remain to
be understood.

In this paper we investigate the effect of a magnetic field directly on the bosonic two leg
ladder. We study the Meissner-vortex transition in this system, and the nature of the
vortex phase. We show that plateaus in the magnetization for commensurate
values of the field exist, in a way similar to the Mott transition in one dimension for commensurate values of the
filing. We analyse the consequences for transport in such bosonic ladders.

The lattice Hamiltonian of the bosonic two leg ladder in a magnetic field is:
\begin{eqnarray}
\label{eq:josephson_ladder_hamiltonian}
H&=&-t_\parallel \sum_{i,p=1,2} (b^\dagger_{i+1,p} {\rm e}^{i e^*
a A_{\parallel,p}(i)} b_{i,p} + b^\dagger_{i,p}{\rm e}^{-i e^*
a A_{\parallel,p}(i)}b_{i+1,p}) \nonumber \\
 &-&t_\perp \sum_i (b^\dagger_{i,2} e^{i
e^* A_\perp(i)} b_{i,1} + b^\dagger_{i,1} e^{-i
e^*A_\perp(i)} b_{i,2})\nonumber \\
& +& U\sum_{i,p} n_{i,p}(n_{i,p}-1) +V n_{i,1}n_{i,2}
\end{eqnarray}
where the density $n_{i,p}=b^\dagger_{i,p} b_{i,p}$
and the magnetic field is introduced via the Peierls
substitution. In addition to their parallel $t_\parallel$ and perpendicular
$t_\perp$ hopping the bosons repel via an on-site $U$ and an inter-chain
$V$ repulsion. The presence of a uniform magnetic field implies that:
\begin{equation}\label{eq:discrete_green}
\oint {\bf A}\cdot d{\bf l}=
A_\perp(i+1)+a A_{\parallel,1}(i)-A_\perp(i)-a A_{\parallel,2}(i)=\Phi
\end{equation}
Where $\Phi$ is the flux of the magnetic field through a plaquette.
The low energy properties of (\ref{eq:josephson_ladder_hamiltonian})
are more transparent\cite{orignac_2chain_bosonic} using a ``bosonized'' representation of the
boson operators \cite{haldane_bosons}. One obtains:
\begin{eqnarray}\label{eq:bosonized_josephson_ladder}
H=\sum_{p=1,2} \int \frac{dx}{2\pi} \left[uK (\pi \Pi_p -e^*
A_{\parallel,p})^2 +
\frac u K (\partial_x \phi_p)^2\right] \nonumber \\
-\frac{t_\perp}{\pi a} \int dx \cos (\theta_1 -\theta_2 +e^*
A_\perp(x)) +\nonumber \\
 \frac{Va}{\pi^2}\int dx \partial_x \phi_1 \partial_x \phi_2 +
\frac{2Va}{(2\pi a)^2} \int dx \cos (2\phi_1 -2\phi_2)
\end{eqnarray}
where $\phi_i$ and $\pi\Pi_i=\nabla_x\theta_i$ are conjugate variables and
the boson annihilation operator is given by:
\begin{equation}\label{eq:bosonize_boson}
\psi_p(x=na)=\frac{b_n}{\sqrt{a}}=\frac{e^{i \theta_p(x)}}{\sqrt{2\pi a}}
\end{equation}
where $a$ is the lattice spacing along the legs of the ladder. $u$ is
the sound velocity of the collective density oscillations.  The linear
nature of the spectrum is the signature that true superfluidity exist
in a single chain even without long range order \cite{mikeska_supra_1d}. $K$ is
the Luttinger parameter, directly related to the compressibility of
the system, and incorporating
all microscopic interaction effects. For a simple onsite repulsion
$1<K<\infty$, 
$K=\infty$ being free bosons and $K=1$ hard core bosons. For more
general (longer range) interactions
all values of $K$ are in principle allowed.
If one introduces the symmetric and antisymmetric combinations
$\phi_{s,a}=\frac{\phi_1\pm\phi_2}{\sqrt{2}}$ (and similar combinations for $\Pi$ and $A$),
Eq.~(\ref{eq:bosonized_josephson_ladder}) reads\cite{orignac_2chain_bosonic}:
\begin{eqnarray}
H &= & H^0_s+H^0_a - \frac{t_\perp}{\pi a} \int dx \cos
(\sqrt{2} \theta_a +e^* A_\perp(x))  \nonumber \\
& + &\frac{2Va}{(2\pi a)^2} \int dx
\cos \sqrt{8} \phi_a
\end{eqnarray}
where ($\nu=s,a$):
\begin{equation}
H^0_\nu = \int \frac{dx}{2\pi} \left[ u_\nu K_\nu (\pi \Pi_\nu - e^* A_\nu)^2 + \frac {u_\nu} {K_\nu}
(\partial_x \phi_\nu)^2\right]
\end{equation}
whith $K_s=K\left(1+\frac{VKa}{\pi u}\right)^{-1/2}$,
$K_a=K\left(1-\frac{VKa}{\pi u}\right)^{-1/2}$.
$A_s$ describes  an Ahronov-Bohm Flux threading the system.
In the following, we
assume that there is no such flux, so $A_s=0$. We
also assume that $K_a>1$, so that the term $\cos \sqrt{8}\phi_a$ is
irrelevant. This
leads to an Hamiltonian similar to the one derived for the
Josephson ladder\cite{kardar_josephson_ladder,granato_josephson_ladder}.
The total and antisymmetric longitudinal current $j_{s,a}=j_1 \pm j_2$, and the
current perpendicular to the ladder are given by ($\nu=s,a$):
\begin{eqnarray}
j_\nu &=& uK e^* \sqrt{2} (\Pi_\nu -\frac {e^*} \pi A_\nu) \\
j_\perp &=& \frac {e^* t_\perp}{\pi a} \sin (\sqrt{2}\theta_a +e^* A_\perp)
\end{eqnarray}

\begin{figure}
\centerline{\includegraphics[width=\figwidth]{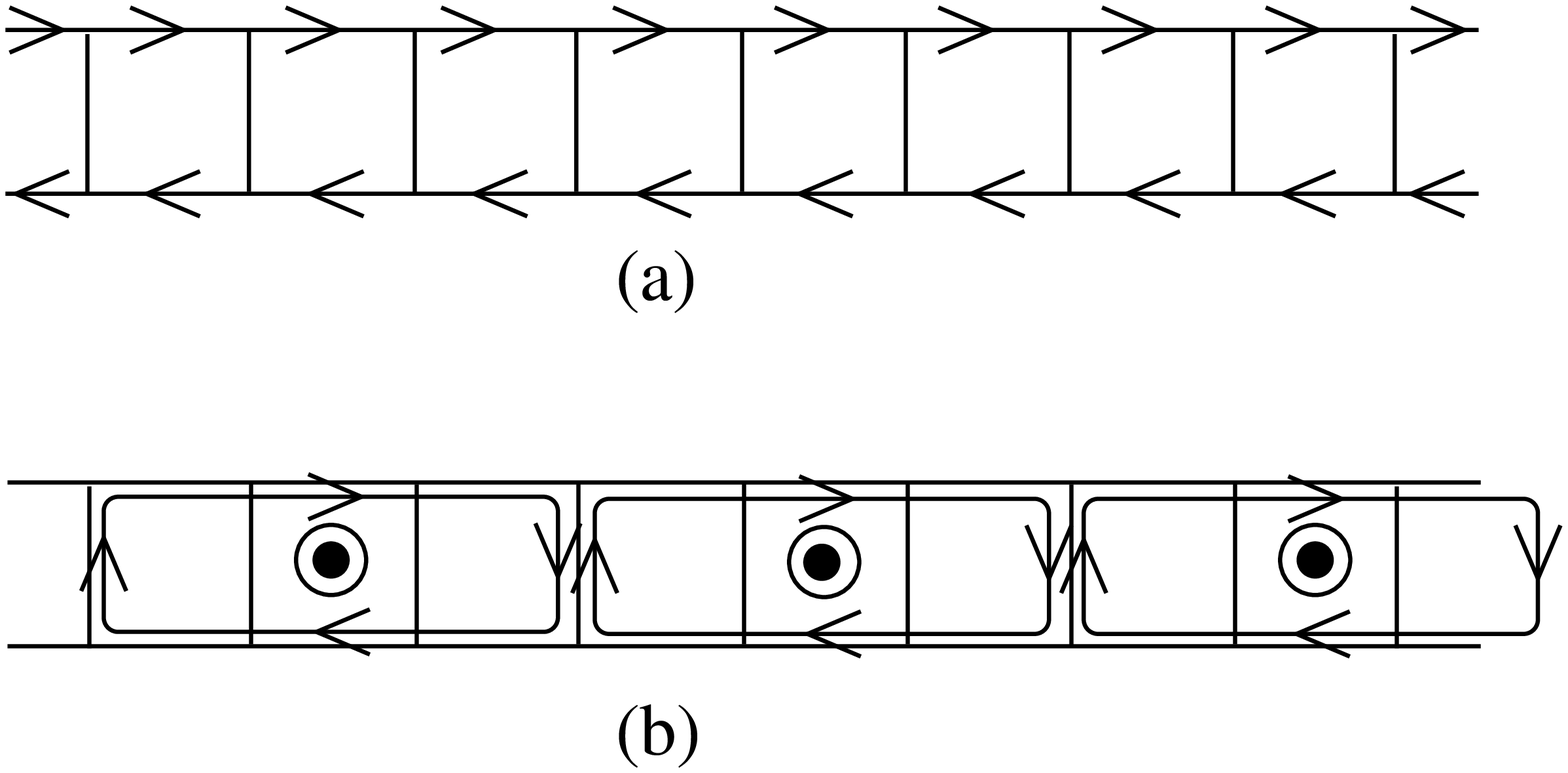}}
\caption{The phases of the bosonic two leg ladder in the presence of a magnetic field.
(a) The low field phases. Current only exist along the legs of the ladder leading to screening
of the applied magnetic field. (b) the high field phase. Current exist both on the legs and the
rungs of the ladder leading to a vortex lattice (here shown with a vortex every three sites).
At commensurable flux the vortex lattice is perfectly
ordered (at $T=0$) whereas away from commensurability the vortex
system has only quasi long range positional order and no perfect crystalline order.}
\label{fig:currents}
\end{figure}

$H_s$ gives a simple dynamics, the behavior of $H_a$ is richer.
In the absence of a magnetic field, $H_a$
is a sine-Gordon Hamiltonian. For $K>1/4$, it develops an order in the field
$\theta_a$ with $\langle \theta_a\rangle=0$ and a gap in the excitation spectrum $\Delta_a \sim
\frac u a \left(\frac{t_\perp a}{u}\right)^{\frac 1 {2-1/(2K)}}$. This
describes the locking of the relative phases by Josephson coupling.
Let us now consider the effect of a magnetic field such that  the flux
per plaquette $\Phi=\frac{2\pi}{e^*}\frac p q$, where $p,q$ are two
mutually prime integers. Using the gauge
$A_a=0$, $A_\perp=\Phi \frac x a$, we can write:
\begin{eqnarray}
\label{eq:Ha_rational_flux}
H_a& = &\int \frac{dx}{2\pi} \left[ u_a K_a (\pi \Pi_a)^2 + \frac
{u_a} {K_a}
(\partial_x \phi_a)^2\right] \nonumber \\ &-& \frac{t_\perp}{\pi a}
\int dx \cos \left(\sqrt{2} \theta_a +\frac{2\pi p}{q} \frac x a\right)
\end{eqnarray}
The Hamiltonian~(\ref{eq:Ha_rational_flux})
is similar to the one describing the
Mott transition in one dimension \cite{giamarchi_mott_shortrev}, where any commensurability
can in principle give a transition depending on $K$. Indeed, even
if the presence of an oscillating phase
in (\ref{eq:Ha_rational_flux}) makes $t_\perp$ naively
irrelevant, perturbation theory to order $q$ in
$\frac{t_\perp a}{\pi u_a}$ shows \cite{giamarchi_curvature,schulz_losalamos}
that a term:
\begin{equation}
\label{eq:commensurate_perturbation}
H_q=\frac{g_q}{\pi a} \int dx \cos q\sqrt{2} \theta_a
\end{equation}
exists with $g_q a /(\pi u_a) \sim (t_\perp a/\pi u_a)^q$ in
perturbation. More generally this term
is allowed
by symmetry and is present in the Hamiltonian even beyond perturbation
theory.
The term (\ref{eq:commensurate_perturbation}) opens a gap $\Delta_{p/q}=\frac{\pi u}{a} \left(\frac
{t_\perp a}{\pi u_a}\right)^{\frac{2K_a q}{4K_a-q^2}}$ provided $K_a>\frac {q^2} 4$.
The expectation value
of $\theta_a$ in the gapped phase is $\langle \theta_a
\rangle=(2k+1)\pi/(q\sqrt{2})$ where $k=0,\ldots,q-1$. This
corresponds to  a $q-$fold degenerate ground state resulting from  the
breaking of the discrete translation symmetry, in direct analogy with
Mott systems\cite{giamarchi_mott_shortrev} or spin
systems\cite{oshikawa,cabra}. The corresponding expectation value of $j_\perp(x)$ is:
\begin{equation}\label{eq:current_rational_flux}
\langle j_\perp(x)\rangle \propto \frac{e^* t_\perp}{\pi a} \sin \left[\frac{(2k+1)
\pi}{q} + \frac{2\pi p} a \frac x a \right]
\end{equation}
giving a periodic pattern of the transverse currents of period
$qa$ as shown in Fig.~\ref{fig:currents}.
This pattern corresponds
to a vortex lattice phase pinned to the microscopic lattice
with $p$ vortices in a supercell of $q$ sites, leading to an
average vortex density $\overline{\rho}_V=\frac p q$.
Let us now take a flux per plaquette which is an
irrational multiple of the quantum of flux.
If this value is close to a rational number with a small enough denominatorwe can decompose
$e^*\Phi=2\pi \frac p q + \delta \Phi$.
The most convenient
gauge choice is then  $A_\perp=(2 \pi e^* p)/ q $ and $A_a=
(\delta \Phi)/(a \sqrt{2})$.
Introducing the field $P_a$ conjugate to $\theta_a$, $\pi
P_a=\partial_x\phi$, $H_a$ becomes:
\begin{eqnarray}
H_a &= &\int \frac{dx}{2\pi} \left[ \frac u K (\pi P_a)^2 +\frac u K
\left(\partial_x \theta_a +\frac{e^* \Phi}{a \sqrt{2}}\right)^2 \right]
\nonumber \\
&- & \frac{g_q}{\pi a} \int dx \cos q\sqrt{2} \theta_a
\end{eqnarray}
The deviation of the flux per plaquette  from a rational value in the
unit of the quantum of flux  causes a
commensurate-incommensurate
transition \cite{japaridze_cic_transition,pokrovsky_talapov_prl,schulz_cic2d}:
At low deviation, the system is in its commensurate phase, $\sqrt{2}\langle \theta_a
\rangle=(2k+1)\pi/q$. This implies $\langle j_a \rangle=-uK
\frac{(e^*)^2 \delta \Phi}{\pi
a}$ and no modification of the perpendicular current $\langle \delta
j_\perp \rangle=0$ and $ \langle \delta j_\perp(x) \delta j_\perp(x')
\rangle \sim e^{-|x-x'|/\xi_a}$ with $\xi_a=u/\Delta_a$.
In particular, for low magnetic field ($p=0,q=1$),
there is a Meissner current circulating
only on the edges of the system (see Fig.~\ref{fig:currents}) and
proportional to the
applied vector potential\cite{kardar_josephson_ladder}.
For a large enough deviation from rationality, $\frac u a \delta \Phi
> \Delta_{p/q}$, there is a transition
to a phase where the $\cos q \sqrt{2}
\theta_a$ operator is irrelevant.
In this phase $\langle \partial_x \theta_a
\rangle/(\pi\sqrt2)$ is non zero an defines an order parameter. In the
small applied field case\cite{kardar_josephson_ladder}, this order
parameter is just the vortex
density $\rho_V$. In the general case, this order parameter measures
the deviation $\delta \rho_V =\rho_V-\frac p q$
of the vortex density from a rational value.
The average transverse current is zero in this phase.
However, transverse currents correlations have an oscillating component that
decays with a power law as
\begin{equation} \label{eq:fluccur}
\langle j_\perp(x) j_\perp(x') \rangle \sim \frac{\cos [2\pi \overline{\rho}_V(x-x')]}{|x-x'|^{1/K_a^*}}
\end{equation}
where $K_a^*$ is the renormalized Luttinger parameter. It would be worthwhile to check in a numerical simulation such as
\cite{nishiyama_josephson_ladder} whether such current correlations are
present.
There is a pattern of transverse currents
alternating along the rungs of the ladder.
One can identify the solitons in $\theta_a$ with vortices
surrounded by circulating currents (see figure \ref{fig:currents}). If  a
vortex is at position $x$, it causes a jump of $2\pi$ in
$\theta_1-\theta_2$ at this point. This enables us to write the density of these one-dimensional
objects as \cite{haldane_bosons}:
\begin{equation} \label{eq:vordens}
\rho_V(x)=\frac{1}{\pi
\sqrt{2}} \partial_x \theta_a + \sum_{m=-\infty}^{m=\infty} \frac{C_m}{\pi a} \
e^{i m [2\pi\overline{\rho}_V x + \sqrt{2} (\theta_a-\langle \theta_a \rangle)]}
\end{equation}
When the density of vortices is $p/q$, the potential
energy term (\ref{eq:commensurate_perturbation}) can be
interpreted as the coupling of the vortex density to a pinning
potential of period $q$ lattice spacings. Therefore we recover the
interpretation in term of pinning of the vortices by the underlying
microscopic lattice and the analogy with the Mott
transition\cite{giamarchi_mott_shortrev} in
$d=1$. Eq.~(\ref{eq:vordens}) gives as vortex current
$j_V(x)=-\frac {\partial_t \theta_a}{\pi \sqrt{2}}$
The application of a voltage $V_1-V_2$ between the
chains  gives: $\langle \partial_x \phi_a \rangle=e^*
\frac{V_1-V_2}{u_a K_a \sqrt{2}}$ and  $V_1-V_2=E_\perp b
=-\frac{2\pi}{e^*} j_V$, which is just Faraday's law. This completes the
identification of the incommensurate phase as an unpinned vortex
lattice. Because of the linearization of the spectrum
in~(\ref{eq:bosonized_josephson_ladder}) there is no Laplace force on
vortices. This pathology could be cured by putting back the band curvature.

\begin{figure}
\centerline{\includegraphics[width=\figwidth]{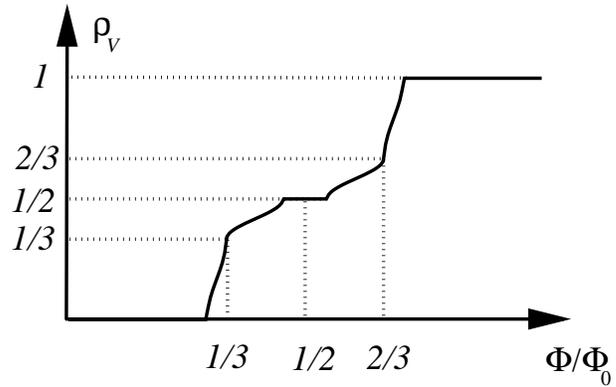}}
\caption{Sketch of the staircase in the magnetization of the
bosonic or Josephson ladder. On the figure, $K_a=4$, so that only the
plateaus obtained for $p/q=0,1/3,1/2,2/3,1$ survive to the quantum
fluctuations. Note that the width of the $1/3$ and $2/3$ plateaus is
already extremely reduced compared with the width of the $1/2$
plateau.}
\label{fig:devil}
\end{figure}

In the classical case
\cite{denniston_classical_ladder}, a vortex lattice
phase is obtained each time the flux per plaquette in a rational
multiple of the quantum of flux, leading to a devil's staircase
structure in the behavior of the magnetization. Here,
the quantum fluctuations wipe out the large fractions for which $q^2>4K_a$,
so only some plateaux remain as shown on Fig.~\ref{fig:devil}.
As quantum fluctuations are
reduced, more and more plateaus in the magnetization curve are
formed. It is interesting to study how the width of a magnetization
plateau increases as a function of $K_a$ or $t_\perp$. Standard RG
calculation of the gap show that the width of a plateau behaves as
$\exp(-C/\sqrt{t_\perp-t_\perp^c})$ close to the threshold.
For $\Phi=\frac{\pi}{e^*}$, a vortex lattice
phase has been obtained in numerical
simulations \cite{nishiyama_josephson_ladder}. It remains to be seen
whether such phase is a Luttinger Liquid or a pinned vortex lattice.

\begin{figure}
\centerline{\includegraphics[angle=-90,width=\figwidth]{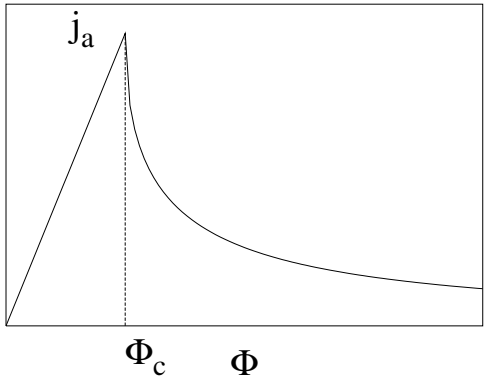}}
\caption{The behavior of the screening current with $\Phi$. For
$\Phi<\Phi_c$, the system is in the Meissner phase and the
screening current increases linearly with $\Phi$. For $\Phi>\Phi_c$,
the screening current decreases with the applied field.}
\label{fig:j_vs_phi}
\end{figure}

Close to the transition
the coefficient $K_a^*$ is universal and $K_a^*=q^2/2$. It is possible in
the limit $\delta \Phi \to \delta\Phi_{c1}(p/q)$ to refermionize
\cite{japaridze_cic_transition,schulz_cic2d} yielding an  average
vortex density $\langle \rho_V \rangle=\frac {e^*}{2 \pi
a}\sqrt{\Phi^2-\Phi_c^2} $  and vortex current $\langle j_a \rangle =
\frac{u (e^*)^2}{2\pi a} (\sqrt{\Phi^2-\Phi_c^2} - \Phi)$.
The transition from the pinned vortex lattice state to the vortex
liquid state is thus continuous.
The behavior of the average current as a function of flux is
plotted on figure \ref{fig:j_vs_phi}.
This is reminiscent of the behavior of a Type II
superconductor, however the growth
of the magnetization is here quite different\cite{kardar_josephson_ladder} from the
standard $B = H+M \sim \frac{1}{\ln^2(H-H_{c1})}$ law.

How could the Meissner and the vortex lattice phase be detected in an
experiment ? First,
$h_{c_1}=\frac{\Phi_c}{ab}$ must be smaller than the spin gap.
Otherwise, the magnetic field would break the Cooper pairs before vortices are
nucleated, a situation  reminiscent of type I
superconductivity. Second, in a standard material,
$a\simeq 1 $\AA so that $h_{c_1}\sim \frac{h}{2e^* a^2}\simeq 2.1\times
10^{5} \mathrm{T}$ making the vortex lattice unobservable.
For Josephson junction ladders, $a\simeq 1 \mu\mathrm{m}$,
giving a typical $h_{c1}\simeq 2.1\times 10^{-3} \mathrm{T}$
which is easily achieved. In principle, the formation of the vortex lattice
in this system could be observed by magnetization measurements or by resolving individual vortices as
was done for other mesoscopic systems\cite{levy_mesoscopic_rings}.
The Meissner transition also affects  transport properties at a
commensurate filling of an integer number of Cooper pairs per island.
In this case, a $\cos 2\phi_s \cos 2\phi_a$ term is present and favors an insulating
state\cite{granato_josephson_ladder}.  The
interchain Josephson coupling
$\cos \sqrt{2}\theta_a$ competes with this term making the system
conducting  when large enough \cite{donohue_commensurate_bosonic_ladder}.
The application of $H > H_{c1}$ suppresses the
interchain Josephson coupling turning the ladder insulating in the
vortex phase. The Meissner transition is thus accompanied by a
superconductor-insulator transition in this case.
At incommensurate filling, the magnetic field still affects transport
properties if there is an artificial defect in the ladder, which adds a
term $V_1 \cos \sqrt{2}\phi_s(x=0) \cos \sqrt{2} \phi_a(x=0) + V_2
\cos \sqrt{8} \phi_s(x=0)$ to the
Hamiltonian. In the Meissner phase, the field $\cos \sqrt{2}\phi_a$ is
disordered and the $V_2$ term dominates transport. If $K>1$
this term is irrelevant and the conductance tends at low temperatures
to the quantum of conductance. In the vortex phase
on the other hand, the field $\cos \sqrt{2} \phi_a$ has
quasi long range order and the $V_1$ term dominates transport,
leading (for $K<2$) to an
insulating state at zero temperature. More generally, even when $K<1$ or $K>2$
the conductance has a qualitatively different temperature dependence in the Meissner and in the vortex phase.
\begin{acknowledgments}
We thank P. Donohue, B. Dou\c{c}ot, A. Georges and J. Vidal for
illuminating discussions, and an a referee for pointing out to
us Ref.~\onlinecite{kardar_josephson_ladder,granato_josephson_ladder}.
This work has been supported in part by the NATO grant No. 971615 .
\end{acknowledgments}


\begin{thebibliography}{24}
\expandafter\ifx\csname natexlab\endcsname\relax\def\natexlab#1{#1}\fi
\expandafter\ifx\csname bibnamefont\endcsname\relax
  \def\bibnamefont#1{#1}\fi
\expandafter\ifx\csname bibfnamefont\endcsname\relax
  \def\bibfnamefont#1{#1}\fi
\expandafter\ifx\csname citenamefont\endcsname\relax
  \def\citenamefont#1{#1}\fi
\expandafter\ifx\csname url\endcsname\relax
  \def\url#1{\texttt{#1}}\fi
\expandafter\ifx\csname urlprefix\endcsname\relax\def\urlprefix{URL }\fi
\providecommand{\bibinfo}[2]{#2}
\providecommand{\eprint}[2][]{\url{#2}}

\bibitem[{\citenamefont{Mikeska and Schmidt}(1970)}]{mikeska_supra_1d}
\bibinfo{author}{\bibfnamefont{H.~J.} \bibnamefont{Mikeska}} \bibnamefont{and}
  \bibinfo{author}{\bibfnamefont{H.}~\bibnamefont{Schmidt}},
  \bibinfo{journal}{J. Low Temp. Phys} \textbf{\bibinfo{volume}{2}},
  \bibinfo{pages}{371} (\bibinfo{year}{1970}).

\bibitem[{\citenamefont{Dagotto and Rice}(1996)}]{dagotto_2ch_review}
\bibinfo{author}{\bibfnamefont{E.}~\bibnamefont{Dagotto}} \bibnamefont{and}
  \bibinfo{author}{\bibfnamefont{T.~M.} \bibnamefont{Rice}},
  \bibinfo{journal}{Science} \textbf{\bibinfo{volume}{271}},
  \bibinfo{pages}{618} (\bibinfo{year}{1996}), \bibinfo{note}{and references
  therein}.

\bibitem[{\citenamefont{Bockrath et~al.}(1999)\citenamefont{Bockrath, Cobden,
  Lu, Rinzler, Smalley, Balents, and Mceuen}}]{bockrath_luttinger_nanotubes}
\bibinfo{author}{\bibfnamefont{M.}~\bibnamefont{Bockrath}},
  \bibinfo{author}{\bibfnamefont{D.~H.} \bibnamefont{Cobden}},
  \bibinfo{author}{\bibfnamefont{J.}~\bibnamefont{Lu}},
  \bibinfo{author}{\bibfnamefont{A.~G.} \bibnamefont{Rinzler}},
  \bibinfo{author}{\bibfnamefont{R.~E.} \bibnamefont{Smalley}},
  \bibinfo{author}{\bibfnamefont{L.}~\bibnamefont{Balents}}, \bibnamefont{and}
  \bibinfo{author}{\bibfnamefont{P.~L.} \bibnamefont{Mceuen}},
  \bibinfo{journal}{Nature} \textbf{\bibinfo{volume}{397}},
  \bibinfo{pages}{598} (\bibinfo{year}{1999}).

\bibitem[{\citenamefont{Fazio and {van der
  Zant}}(2000)}]{fazio_josephson_review}
\bibinfo{author}{\bibfnamefont{R.}~\bibnamefont{Fazio}} \bibnamefont{and}
  \bibinfo{author}{\bibfnamefont{H.}~\bibnamefont{{van der Zant}}},
  \emph{\bibinfo{title}{Quantum phase transitions and vortex dynamics in
  superconducting networks}} (\bibinfo{year}{2000}),
  \bibinfo{note}{cond-mat/0011152}.

\bibitem[{\citenamefont{{van Oudenaarden} and
  Mooij}(1996)}]{vanoudenaarden_josephson_mott}
\bibinfo{author}{\bibfnamefont{A.}~\bibnamefont{{van Oudenaarden}}}
  \bibnamefont{and} \bibinfo{author}{\bibfnamefont{J.~E.} \bibnamefont{Mooij}},
  \bibinfo{journal}{Phys. Rev. Lett.} \textbf{\bibinfo{volume}{76}},
  \bibinfo{pages}{4947} (\bibinfo{year}{1996}).

\bibitem[{\citenamefont{{van Oudenaarden} et~al.}(1996)\citenamefont{{van
  Oudenaarden}, V{\'a}rdy, and Mooij}}]{vanoudenaarden_josephson_localization}
\bibinfo{author}{\bibfnamefont{A.}~\bibnamefont{{van Oudenaarden}}},
  \bibinfo{author}{\bibfnamefont{S.~. J.~K.} \bibnamefont{V{\'a}rdy}},
  \bibnamefont{and} \bibinfo{author}{\bibfnamefont{J.}~\bibnamefont{Mooij}},
  \bibinfo{journal}{Phys. Rev. Lett.} \textbf{\bibinfo{volume}{77}},
  \bibinfo{pages}{4257} (\bibinfo{year}{1996}).

\bibitem[{\citenamefont{Bradley and Doniach}(1984)}]{bradley_josephson_chain}
\bibinfo{author}{\bibfnamefont{R.~M.} \bibnamefont{Bradley}} \bibnamefont{and}
  \bibinfo{author}{\bibfnamefont{S.}~\bibnamefont{Doniach}},
  \bibinfo{journal}{Phys. Rev. B} \textbf{\bibinfo{volume}{30}},
  \bibinfo{pages}{1138} (\bibinfo{year}{1984}).

\bibitem[{\citenamefont{Glazman and Larkin}(1997)}]{glazman_josephson_1d}
\bibinfo{author}{\bibfnamefont{L.}~\bibnamefont{Glazman}} \bibnamefont{and}
  \bibinfo{author}{\bibfnamefont{A.}~\bibnamefont{Larkin}},
  \bibinfo{journal}{Phys. Rev. Lett.} \textbf{\bibinfo{volume}{79}},
  \bibinfo{pages}{3736} (\bibinfo{year}{1997}), \eprint{cond-mat/9705169}.

\bibitem[{\citenamefont{Denniston and Tang}(1995)}]{denniston_classical_ladder}
\bibinfo{author}{\bibfnamefont{C.}~\bibnamefont{Denniston}} \bibnamefont{and}
  \bibinfo{author}{\bibfnamefont{C.}~\bibnamefont{Tang}},
  \bibinfo{journal}{Phys. Rev. Lett.} \textbf{\bibinfo{volume}{75}},
  \bibinfo{pages}{3930} (\bibinfo{year}{1995}).

\bibitem[{\citenamefont{Kardar}(1986)}]{kardar_josephson_ladder}
\bibinfo{author}{\bibfnamefont{M.}~\bibnamefont{Kardar}},
  \bibinfo{journal}{Phys. Rev. B} \textbf{\bibinfo{volume}{33}},
  \bibinfo{pages}{3125} (\bibinfo{year}{1986}).

\bibitem[{\citenamefont{Granato}(1990)}]{granato_josephson_ladder}
\bibinfo{author}{\bibfnamefont{E.}~\bibnamefont{Granato}},
  \bibinfo{journal}{Phys. Rev. B} \textbf{\bibinfo{volume}{42}},
  \bibinfo{pages}{4797} (\bibinfo{year}{1990}).

\bibitem[{\citenamefont{Nishiyama}(2000)}]{nishiyama_josephson_ladder}
\bibinfo{author}{\bibfnamefont{Y.}~\bibnamefont{Nishiyama}},
  \bibinfo{journal}{Eur. Phys. J. B} \textbf{\bibinfo{volume}{17}},
  \bibinfo{pages}{295} (\bibinfo{year}{2000}), \eprint{cond-mat/0006311}.

\bibitem[{\citenamefont{Orignac and Giamarchi}(1998)}]{orignac_2chain_bosonic}
\bibinfo{author}{\bibfnamefont{E.}~\bibnamefont{Orignac}} \bibnamefont{and}
  \bibinfo{author}{\bibfnamefont{T.}~\bibnamefont{Giamarchi}},
  \bibinfo{journal}{Phys. Rev. B} \textbf{\bibinfo{volume}{57}},
  \bibinfo{pages}{11713} (\bibinfo{year}{1998}).

\bibitem[{\citenamefont{Haldane}(1981)}]{haldane_bosons}
\bibinfo{author}{\bibfnamefont{F.~D.~M.} \bibnamefont{Haldane}},
  \bibinfo{journal}{Phys. Rev. Lett.} \textbf{\bibinfo{volume}{47}},
  \bibinfo{pages}{1840} (\bibinfo{year}{1981}).

\bibitem[{\citenamefont{Giamarchi}(1997)}]{giamarchi_mott_shortrev}
\bibinfo{author}{\bibfnamefont{T.}~\bibnamefont{Giamarchi}},
  \bibinfo{journal}{Physica B} \textbf{\bibinfo{volume}{230-232}},
  \bibinfo{pages}{975} (\bibinfo{year}{1997}).

\bibitem[{\citenamefont{Giamarchi and Millis}(1992)}]{giamarchi_curvature}
\bibinfo{author}{\bibfnamefont{T.}~\bibnamefont{Giamarchi}} \bibnamefont{and}
  \bibinfo{author}{\bibfnamefont{A.~J.} \bibnamefont{Millis}},
  \bibinfo{journal}{Phys. Rev. B} \textbf{\bibinfo{volume}{46}},
  \bibinfo{pages}{9325} (\bibinfo{year}{1992}).

\bibitem[{\citenamefont{Schulz}(1994)}]{schulz_losalamos}
\bibinfo{author}{\bibfnamefont{H.~J.} \bibnamefont{Schulz}},
  \emph{\bibinfo{title}{Strongly Correlated Electronic Materials: The Los
  Alamos Symposium 1993}} (\bibinfo{publisher}{Addison--Wesley},
  \bibinfo{address}{Reading, Massachusetts}, \bibinfo{year}{1994}), p.
  \bibinfo{pages}{187}.

\bibitem[{\citenamefont{Oshikawa et~al.}(1997)\citenamefont{Oshikawa, Yamanaka,
  and Affleck}}]{oshikawa}
\bibinfo{author}{\bibfnamefont{M.}~\bibnamefont{Oshikawa}},
  \bibinfo{author}{\bibfnamefont{M.}~\bibnamefont{Yamanaka}}, \bibnamefont{and}
  \bibinfo{author}{\bibfnamefont{I.}~\bibnamefont{Affleck}},
  \bibinfo{journal}{Phys. Rev. Lett.} \textbf{\bibinfo{volume}{78}},
  \bibinfo{pages}{1984} (\bibinfo{year}{1997}).

\bibitem[{\citenamefont{Cabra et~al.}(1997)\citenamefont{Cabra, Honecker, and
  Pujol}}]{cabra}
\bibinfo{author}{\bibfnamefont{D.}~\bibnamefont{Cabra}},
  \bibinfo{author}{\bibfnamefont{A.}~\bibnamefont{Honecker}}, \bibnamefont{and}
  \bibinfo{author}{\bibfnamefont{P.}~\bibnamefont{Pujol}},
  \bibinfo{journal}{Phys. Rev. Lett.} \textbf{\bibinfo{volume}{79}},
  \bibinfo{pages}{5126} (\bibinfo{year}{1997}).

\bibitem[{\citenamefont{Japaridze and
  Nersesyan}(1978)}]{japaridze_cic_transition}
\bibinfo{author}{\bibfnamefont{G.~I.} \bibnamefont{Japaridze}}
  \bibnamefont{and} \bibinfo{author}{\bibfnamefont{A.~A.}
  \bibnamefont{Nersesyan}}, \bibinfo{journal}{JETP Lett.}
  \textbf{\bibinfo{volume}{27}}, \bibinfo{pages}{334} (\bibinfo{year}{1978}).

\bibitem[{\citenamefont{Pokrovsky and Talapov}(1979)}]{pokrovsky_talapov_prl}
\bibinfo{author}{\bibfnamefont{V.~L.} \bibnamefont{Pokrovsky}}
  \bibnamefont{and} \bibinfo{author}{\bibfnamefont{A.~L.}
  \bibnamefont{Talapov}}, \bibinfo{journal}{Phys. Rev. Lett.}
  \textbf{\bibinfo{volume}{42}}, \bibinfo{pages}{65} (\bibinfo{year}{1979}).

\bibitem[{\citenamefont{Schulz}(1980)}]{schulz_cic2d}
\bibinfo{author}{\bibfnamefont{H.~J.} \bibnamefont{Schulz}},
  \bibinfo{journal}{Phys. Rev. B} \textbf{\bibinfo{volume}{22}},
  \bibinfo{pages}{5274} (\bibinfo{year}{1980}).

\bibitem[{\citenamefont{L{\'e}vy et~al.}(1990)\citenamefont{L{\'e}vy, Dolan,
  Dunsmuir, and Bouchiat}}]{levy_mesoscopic_rings}
\bibinfo{author}{\bibfnamefont{L.~P.} \bibnamefont{L{\'e}vy}},
  \bibinfo{author}{\bibfnamefont{G.}~\bibnamefont{Dolan}},
  \bibinfo{author}{\bibfnamefont{J.}~\bibnamefont{Dunsmuir}}, \bibnamefont{and}
  \bibinfo{author}{\bibfnamefont{H.}~\bibnamefont{Bouchiat}},
  \bibinfo{journal}{Phys. Rev. Lett.} \textbf{\bibinfo{volume}{64}},
  \bibinfo{pages}{2074} (\bibinfo{year}{1990}).

\bibitem[{\citenamefont{Donohue and
  Giamarchi}(2001)}]{donohue_commensurate_bosonic_ladder}
\bibinfo{author}{\bibfnamefont{P.}~\bibnamefont{Donohue}} \bibnamefont{and}
  \bibinfo{author}{\bibfnamefont{T.}~\bibnamefont{Giamarchi}},
  \bibinfo{journal}{Phys. Rev. B} \textbf{\bibinfo{volume}{63}},
  \bibinfo{pages}{180508(R)} (\bibinfo{year}{2001}).

\end{thebibliography}

\end{document}